
\magnification=\magstep1
\def\newline{\hfil\break}
\def\dul{\rlap{\raise 6.75pt\hbox{\hskip 4pt$\cdot$}}}
\def\undersim{\ut}
\def\lappreq{\undersim{<}}

\def\lappreq{\undersim{<}}
\baselineskip=12pt
\def\oneskip{\vskip\baselineskip}         
\font\bf=cmbx10          
\hoffset=1.0truecm       
\voffset=0.15truecm       
\hsize=16.truecm       
\vsize=21.0truecm        
\centerline{\null}     
$~$
\vskip 0.85truecm
\centerline{\bf The evolution of the globular cluster system in a
triaxial galaxy:}
\centerline{\bf can a galactic nucleus form by globular cluster
capture?}
\oneskip
\centerline {Roberto Capuzzo--Dolcetta $^1$}
\vfill\noindent
\oneskip\noindent
\oneskip\noindent
\bigskip\noindent
---------------------
\newline
$^1~$ Istituto Astronomico Universit\` a \lq La Sapienza\rq ,
via G.M. Lancisi 29, I-00161, Roma, Italy
\medskip\noindent
\par\eject
\centerline {ABSTRACT}
\newline Among the possible phenomena inducing
evolution of the globular cluster system in an elliptical galaxy,
dynamical friction due to field stars and tidal disruption caused
by a central nucleus are of crucial importance.
The aim of this paper is the study of the
evolution of the globular cluster system in a triaxial galaxy in
presence of these phenomena. In particular, it is examined
the possibility that some of
galactic nuclei have been formed by frictionally decayied globular clusters
moving in a triaxial potential. We find that the initial rapid
growth of the nucleus, due mainly to massive clusters on box
orbits falling in a short time scale into the galactic centre,
is later slowed by tidal disruption induced by
the nucleus itself on less massive clusters in the way
described by Ostriker, Binney \& Saha.
The efficiency of dynamical friction is such to carry
to the centre of the galaxy enough globular cluster mass
available to form a compact nucleus, but the actual modes and results
of cluster--cluster encounters in the central potential well
remain to be investigated.
The mass of the resulting nucleus is determined by the mutual
feedback of the described processes, together with the knowlegde
of the initial spatial, velocity and mass distributions of the globular
cluster family. The effect on the system mass function is studied,
showing the development of a low-- and a high--mass turn over even
with an initially flat mass function.
Moreover, in this paper is discussed the
possibility that the globular cluster fall to the galactic centre
has been cause of primordial violent galactic activity.
An application of the model to M31 is presented.
\newline {\it Subject headings:} clusters: globular--galaxies:
internal motions--
galaxies: evolution--galaxies: nuclei--galaxies: individual (M 31)--quasars
\vfill\par\eject\noindent
\centerline{1. INTRODUCTION}
\medskip\noindent
There is growing evidence of the existence of compact objects in the centre
of galaxies (M31, see Dressler \& Richstone 1988, Kormendy 1988a; M 32,
Tonry 1984, 1987; NGC 4594, Kormendy 1988b; NGC 3115, Kormendy \& Richstone
1992; etc.).
The possibility that the nuclei of some galaxies may consist of
orbitally decayed globular clusters
has been investigated first by Tremaine, Ostriker \& Spitzer
(1975). In this latter paper, they showed that in a spherical galaxy modeled as
a superposition of two isothermal spheres representing the nuclear and
the bulge component of M31, dynamical friction (taken into account
through the standard Chandrasekhar 1943 term) acting
on globular clusters is able to cause the formation of a nucleus
with mass in the range $10^7\div 10^8 M_\odot$, in acceptable fit with
recent estimates
done by Dressler \& Richstone (1988) and by Kormendy (1988a) on the basis of
well defined rotation curves in the inner arcsecs of M 31.
The Tremaine et al. result has been criticised (see Surdin and
Charikov 1988, van den Bergh 1991) because the average metallicity measured
in the inner core seems far too high compared to the M 31 globular
metallicity; indeed, van den Bergh (1969) showed that more than
$97\%$ of the globular clusters in M 31 have integrated spectra
metal poorer than the semistellar nucleus of the parent galaxy.
Anyway, this only means that the dominant nuclear stellar population
in M 31 is not similar (in metallicity) to the {\it presently}
observed globular cluster population, but does not rule out
the possibility that the most massive clusters, already dragged
to the galaxy centre, were significantly metal richer. The latter
is not an entirely ad hoc assumption, because it works for galaxies,
which show an increasing metallicity--mass relation.
Another controversial point is the disruptive effect played by
the nucleus itself on passing--by cluster, which was not examined
in the Tremaine et al. paper, and that should go in the direction
of limiting the nuclear mass growth. Moreover, it is quite
accepted that the bulge component of M 31 is triaxial in shape.
This likely implies
that a significant fraction of globular clusters in this
galaxy --which have been reasonably initially biased toward
radially pointed orbits-- are moving on box--type orbits, for
which good estimates of both dynamical friction (Pesce,
Capuzzo--Dolcetta \& Vietri 1992, hereafter PCV) and tidal disruption
(Ostriker, Binney \& Saha 1989, hereafter OBS) time scales are available
and show significant differences from both the spherical
and axisymmetric cases. In this paper we will study the
time evolution of a family of globular cluster of given
initial mass function (IMF) and phase--space distribution
function (DF) in a self--consistent triaxial galactic model,
with the aim to evaluate the possibility of nucleus formation
via  frictionally decelerated clusters
and to test the hypothesis, raised by OBS, that such kind
of nucleus formation is a self--limiting process, due to the
increasing importance of tidal shredding (Sects. 2--4). It is necessary to
point
out here that a massive central object can act as a regulator of
its mass not only through the destructive tidal action on incoming
clusters but also because (at least when the central
condensation is hard enough)
it tends to break up the region of phase space formerly occupied
by regular boxes (Gerhard \& Binney 1985). Anyway, Gerhard
\& Binney also shown that it is possible to have triaxial systems supporting
box orbits at all energies, even in presence of a density cusp.
This means that a realistic inclusion of the effect of the growing cusp
on the surrounding orbital structure depends on the model, suggesting
us to neglect it here for the sake of simplicity.
While a detailed analysis of the results of cluster--cluster
merging and of the evolutionary density and velocity dispersion
distributions of globular clusters of different masses as compared
to observational data are postponed to a forthcoming paper, here
an application of the model to M 31 is given (see Sect. 5).
It shows how dynamical friction may have rapidly fed an existing nucleus
in this galaxy (and possibly others)
in such a way to produce a luminosity burst lasting
$500~Myr$, i.e. resembling an AGN.
\bigskip\centerline
{2. DYNAMICAL FRICTION AND TIDAL DISRUPTION DECAY TIMES}
\medskip\noindent
\centerline {\it 2.1 Dynamical Friction in a Triaxial Galaxy}
\medskip\noindent
In PCV we have checked the importance of dynamical friction in
determining the orbital evolution of a globular cluster moving
in a triaxial potential. We found that clusters moving on box
orbits feel a significantly larger decelerating effect than
on loop orbits of same energy.
All PCV results rely on numerical integrations of cluster orbits
in the non--rotating, self--consistent triaxial model proposed
by Schwarzschild (1979), whose potential can be expressed
(see de Zeeuw \& Merritt 1983) through the sum of one monopole
component deriving from the density:
$$\rho(r) = {\rho_0\over \left[1+({r\over r_c})^2\right]^{3/2}} \eqno (1)$$
which, projected, corresponds to the modified Hubble law,
and of two spherical harmonics.
For the velocity dispersion we have adopted the data
corresponding to the zero mean--motion self consistent solution of
the Schwarzschild ellipsoid as given by Merritt (1980).
The constants in the model
potential are such that $r_c=200$ pc, $\rho_0 r_c^3=3\times 10^9 M_\odot$;
the axial ratios are 2:1.25:1 (note that these axial ratios are
the same determined by Stark (1977) for the spheroidal component
of M 31).
Orbital energies per unit mass, $E$, are expressed in units of the total
potential depth $4\pi G\rho_0 r_c^2$.
The Schwarzschild model has the appealing characteristic to be
self consistent and to reproduce satisfactorily the typical elliptical
galaxy luminosity profile, so to be considered as a good representation
of triaxial galaxies of moderate axial ratios. Taking also advantage of
one of the results of PCV (i.e. that the dynamical friction times do
not depend on the fine structure of the parent galaxy;
a good modelization of its
role along a box orbit needs only the knowledge of some functions
of the energy alone as well as of the monopole component of the potential)
we can state that the results presented in this paper are quite
representative for a typical triaxial galaxy and can be, in particular, worthly
applied to M 31.
For the purposes of this paper some
orbital evolutions complementig those presented in PCV have been computed.
I limited myself to orbits in the (x,y) plane; non--planar orbits
decay more slowly than planar, but, as we showed in PCV,
the ratio of energy halving times does not exceed $2.5$.
I have computed a new set of box orbits and, for any given energy,
1 quasi--circular and 3 loop orbits of increasing
elongation. The initial conditions of the loop orbits were
chosen in a way to be \lq similar in size\rq ~ to the box orbit
of same energy, i.e. such to have a time--averaged value of $r=\sqrt{x^2+y^2}$
comparable to that of the box orbit.
The initial conditions of the orbits and their total--energy
and halving--energy decay times ($\tau_{df}$ and $\tau_{1/2}$)
are given in Table 1. Figure 1 shows the ratio between the energy
decay times of box to quasi--circular (loop c) and elongated
(loop 3) loop orbits
 and the ratio $\tau_{1/2}/\tau_{df}$ as function of the
initial orbital energy.
The ratio of the typical energy decay
time scales (box/loop) decreases with orbital energy, due to
the peculiarity of box orbits of being always able to reach
the inner region of the galaxy, contrarily to loop orbits
wich are more and more --at increasing orbital energy--
confined in external low density zones.
$\tau_{1/2}$ tends to $\tau_{df}$
at increasing energy because at high energies the most of the time
is spent in carrying the cluster into the galactic core,
where the following decay is rapid.
We note that at high energies the dependence on the azimuthal
action $J$ of the ratio between the decay times of the box
to the loop orbits is steeper
due to the fact that at increasing $E$ box orbits tend to spend a
greater fraction of time out of the core than do elongated loops
(which have low $J$).
Fitting approximations to energy decay times are given in Appendix A.
\medskip\noindent
\centerline {\it 2.2 Impulsive Tidal Disruption}
\medskip\noindent
\par The tidal shocks induced by a galactic nucleus
on globular clusters moving on box orbits
have been studied by OBS. They found that
(in the impulse approximation) the mean time to a devastating
encounter (corresponding to an injection of energy $\delta E\ge GM^2/(5R_h)$,
with $R_h$ half--mass radius) is
(OBS eq. (13)):
$$\tau _{tid}={1\over \mu\pi\sqrt{5}}\sqrt{{GM\over R_h}}{v_n\over v_c^2}
{A_w\over r_cR_h}T_r\eqno (2)$$
where: $\mu\equiv GM_n/(r_cv_c^2)$, $M_n$ is mass of the nucleus, $r_c$
is the galactic core, $v_c$ is the circular velocity at large $r$, $v_n$
is an estimate of the speed of the cluster as it passes at a distance to
the galactic center $r_n$ which is the "boundary" between orbits
behaving as in an
isolated copy of the ellipsoid's potential ($r>r_n$) and keplerian
orbits ($r<r_n$); $A_w$ is the orbit waist's area,
$T_r$ is the half--period of oscillation parallel to the potential
long axis.
Equation (2) states the dependence
of the tidal disruption time on $\sqrt{W_M}/R_h$,
($W_M$ is the gravitational
binding energy per unit mass $W_M\propto {GM\over R_h}$), or,
equivalently, on $\sqrt {\bar{\rho}}$, where $\bar{\rho}$
is the average mass density of the cluster.
Introducing the cluster crossing time
$$T_{cr}\equiv {2R_h\over \sqrt{GM\over R_h}}\eqno(3),$$
equation (3) can be written as:
$$\tau _{tid}={2\over \mu\pi\sqrt{5}}{v_n\over v_c^2}
{A_w\over r_c}{T_r\over T_{cr}}.\eqno(4)$$
In the context of the impulse approximation values too large of
the ratio $T_r/T_{cr}$ in equation (4) are not allowed.
\medskip\noindent\centerline
{2.3 Dynamical Friction and Tidal Disruption as Competitive Effects}
\medskip\noindent
Dynamical friction and tidal disruption time scales are,
in a given potential, both functions of the satellite mass and orbital
energy in different ways. They are both increasing functions of $E$
(the dependence of $\tau_{df}$ on $E$ being steeper) but their dependence
on the globular cluster mass is opposite: indeed $\tau_{tid}$
(see eq. (2)) depends on the square root of the cluster's
mass (for fixed $R_h$) while $\tau_{df}$ inversely scales with $M$.
The ratio $\tau_{tid}/\tau_{df}$ is so an increasing function of
$E$ and $J$ scaled with $(M/R_h)^{3/2}/M_n$.
An acceptable fit to $\tau_{tid}/\tau_{df}$ in the case of box orbits is
$${\tau_{tid}\over \tau_{df}}=(2.13E^2+0.085E){1\over M_n}
\left({M\over R_h}\right)^{3\over 2},$$
with masses in $M_\odot$ and $R_h$ in $pc$.
The $\tau_{tid}/\tau_{df}$ ratio in galaxies with an initially light
nucleus is so large for any reasonable globular clusters mass
to cause a possible merging to the centre causing $\tau_{tid}/\tau_{df}$
to decrease until it is sufficiently small to mean that tidal shredding
is now overwhelming the possibility to further increase the nucleus mass.
As an example $E=0.5$ clusters with mass $5\times 10^5 M_\odot$
and $R_h=1$ pc
(typical of galactic globulars) should stop to merge at center when $M_n$
reaches $2\times 10^8 M_\odot$.
Therefore, the possibility --that we are going to check
in a more detailed way--
that most massive clusters decay rapidly to
the centre of a triaxial galaxy to form an object
of fastly growing mass, stands on robust quantitative basis.
\par If $M_{cl}$ is the sum of the masses of the clusters centrally
decayed at time $t$, and $r_c=200~pc$ and $M_c=3\times 10^9~M_\odot$
are the initial galactic core radius and
mass,  one can think that an idealized result of cluster merging
could be a central supercluster containing $M_{cl}/2$ (distributed
in stars having velocities less than $1/10$ of that of
the bulge stars ) packed in a sphere of radius
$~R_h$, so that the central galactic density is increased up to
$\rho(t)/\rho_0\approx (M_{cl}/M_c)(r_c/R_h)^3$. Assuming $r_c/R_h=100$,
a merging of $300$ clusters of $10^6~M_\odot$ suffices to
enhance the central density of a factor $10^5$, i.e. to
$\sim 3\times 10^7~M_\odot~pc^{-3}$. Actually, an evolution of $R_h$
is expected, due to the injection of energy caused by the
different efficient mechanisms active (dissipation of part of the cluster
center of mass orbital energy into internal random motions and
acquisition of energy from the dynamically hot bulge stars) and
subsequent readjustment to a new equilibrium in the pertinent relaxation
time--scale. Anyway, the situation is not simple to schematize,
due to the fact that the relaxation time scale is also the scale on
which lighter stars escape taking away the acquired excess energy, so
inducing a mass segregation which has a feedback on the core contraction.
The detailed
study of these effects is out of the purposes of this work;
in any case, a part from violent relaxation effects,
the mentioned transient density can be reached if the process of merging is
rapid enough with respect to the two--body relaxation time
of the system. Actually, indicating with $\sigma_1$ and
$\sigma_2$ the one--dimensional velocity dispersions of the bulge
and cluster stars respectively, an estimate of the relaxation time of cluster
stars (Chandrasekhar 1943, eq. (2.377)):
$$T_{rel}={9\sqrt 2\over 256\sqrt{\pi}G^2 log_e\Lambda}{1\over \rho_1}
{\sigma_1\over \sigma_2^2}\eqno(5)$$
leads (letting $\sigma_1=200~Km/sec$, $\sigma_2=20~Km/sec$)
to $\approx 10^{19} yr$ (with a value of 18 for the Coulomb logarithm),
which seems confidently large.
The average interstellar distance
at density $3\times 10^7~M_\odot~pc^{-3}$ is $0.003~pc$, which can
be considered as sufficiently small to ensure the existence of a
significant fraction of stars which are actually encountering
at so short separations, to dissipate tidally part of the orbital energy.
The ratio of physical collision time to relaxation time should
be $t_{coll}/t_{rel}\lappreq 300$ to allow collisions to have influence on core
collapse of the system (Binney \& Tremaine 1987); $t_{coll}/t_{rel}$ is
a steeply decreasing function of $\sigma$ and takes the value $300$
at $\sigma\simeq 65$ Km/sec for solar type stars
$3\times 10^6~M_\odot~pc^{-3}$ and $log_e \Lambda=18$. Actually,
in globular clusters, $\sigma$ (initially $\sim 10$ Km/sec) is expected
to increase significantly during late stages of core collapse
(a factor four as the core mass falls by a factor $10^4$);
an even more relevant heating is expected in our case, due to the
quite violent (occurring on the dynamical friction time scale)
process of merging of clusters in the environment of high velocity bulge
stars.
The qualitative result obtained on the basis of the previous considerations
is that it is likely that a significant fraction of the
merged mass goes to a supercompact configuration, on a time scale
which is anyway not easy to evaluate correctly.
\par Indicating with $\alpha$ the efficiency of the process to
convert the stellar mass carried to the centre by dynamical braking into
a "nucleus" ($\dul M_n=\alpha \dul M_{df}$) , if we let (for simplicity)
$\alpha=1$ we can obtain the value of the nucleus mass needed to inhibit
the process of accretion at the galactic centre
simply by equating expressions (2) and (A2), obtaining:
$$M_n=8.68\times 10^{-15}{v_nA_wT_r\over R_h^{3/2}}
M^{3/2}(1-E)^210^{-g(E,J)}.\eqno(6)$$
The function {\it g(E,J)} is defined in Appendix A;
$M_n$ is in solar masses, provided $v_n$ in km/sec, $A_w$ in
$pc^2$, $T_r$ in $yr$, $R_h$ in $pc$ and $M$ in $M_\odot$.
A representative value of the energy of orbits
in the model used in this paper is a mass--average energy
$$\bar {E}\equiv {\int_0^{r_{max}} \left[{1\over 2} (\sigma_1^2+\sigma_2^2
+\sigma_3^2)+\phi(r)\right]dm\over \int_0^{r_{max}} dm},\eqno(7)$$
where $\sigma_i$ are the principal velocity dispersions
of the Merritt (1980) non rotating model and $\phi(r)$
is the potential generated by the density given by equation (1).
This value would be representative for globular clusters if they had the
same DF of field stars, which is not ensured (see Sect. 3.1).
With $r_{max}=16r_c$ (which is roughly the region of validity
of the Merritt's model), the mean one--dimensional velocity dispersion
is $\sigma_{typ}=\sqrt{\bar{\sigma}/3}\approx 290 Km/sec$
and $\bar E\approx 0.6$.
The evaluation of $M_n$ for box orbits of energy $\bar {E}$ gives
$M_n=2.88 \left({M\over R_h}\right)^{3\over 2}$,
(i.e. $\approx 10^9 M_\odot$ for $M=5\times 10^5 M_\odot$)
reduced to $M_n=0.70 \left({M\over R_h}\right)^{3\over 2}$
for quasi--circular loops of same energy. A factor $2$ of energy
reduction implies a reduction of a factor $7.3$ and $2.6$ for
$M_n$ in the cases of pure box and quasi--circular loops,
respectively.
\newline Figure 2 shows the behaviour of the $M_n$ vs. $M$ relation:
thinking of $M$ as a typical globular cluster mass, the corresponding
$M_n$ represents the typical expected value of the mass of clusters
merged in the centre of the galaxy, which is possibly in the form
of a compact nucleus. If all the clusters have $M=5\times 10^5~M_\odot$
and $E=\bar E$, dynamical friction would require $\approx 1.7~Gyr$
to merge $\approx 3\times 10^8~M_\odot$ at the centre before tidal
shredding would act significantly.
For fixed cluster mass, at
increasing $E$ the possible mass accreted by dynamical
friction increases, as well as the time needed to do that,
If the bulk of orbital energies of
globular clusters is in the range $\left[\bar E/4,\bar E\right]$,
we can see from Figure 2 that an abundant enough family
of $10^6~M_\odot$ clusters would form a nucleus of mass
in the range $\left[7.5\times 10^7~M_\odot,2\times 10^9~M_\odot\right]$ in
the interval of time $\left[1~Gyr,6.5~Gyr\right]$.
\par Equations (2) and (A2) allow to define two limiting masses,
$M_{l,tid}(E,t)$ and
\newline $M_{l,df}(E,J,t)$ that constrain
($M_{tid}\leq M \leq M_{df}$) the
value of the mass of clusters survived up to time $t$ and
having initial energy and action $E$ and $J$. As we can see
from Figure 3 (drawn in the simplifying, and likely, hipothesis that clusters
were all on box orbits), at a given nuclear mass
(here $M_n=3\times 10^7 M_\odot$)
the initial region
of phase space allowed to presently visible clusters is getting
narrow and narrow in time. The intersection between the tidal and
dynamical friction curves defines the minimum energy $E_{min}(t,M_n)$
of clusters still visible at age $t$. The cluster mass corresponding
to $E_{min}$ should be the peak value of the cluster mass distribution
at that age, in what it corresponds to the largest interval of orbital
energies allowed.
It is interesting to note
that at ages around the galactic age ($\approx 15$ Gyr) this mass
is $\approx 2\times 10^5 M_\odot$, very similar to the mass
($10^5 M_\odot$, assuming
$(M/L)_{\odot V}=1.6$) corresponding to the
peak value $-7.1$ of the $M_V$ distribution of
both galactic and M31 globular clusters (van den Bergh 1985).
This mass peak value rises to $10^6 M_\odot$ for $M_n\simeq 10^9 M_\odot$.
Consequently,
the efficiency of the process of globular cluster decay and
possible nucleus formation strongly depends on both the
initial mass function and phase space distribution function.
\par These results, although based on crude approximations,
suggests that the phenomenon of globular merging and its subsequent
regulation may have been active on a short time scale, so that
a further deeper investigation is worth pursuing.
\bigskip
\noindent
\centerline {3. THE EVOLUTIONARY MODEL}
\medskip\noindent
Let us define a "global" distribution function $F(M,\vec x,\vec v,t)$ such that
\newline $dN=F(M,\vec x,\vec v,t)dMd^3\vec xd^3\vec v$ is the number of
clusters
at time $t$ with mass, position and velocity in the ranges $[M,M+dM]$,
$[\vec x,\vec {x}+d\vec x]$, $[\vec v,\vec {v}+d\vec v]$, respectively.
Under a particular choice for the initial distribution
$F(M,\vec x,\vec v,0)\equiv F_0(M,\vec x,\vec v)$, (hereafter index 0 refers
to initial quantities) we can follow the evolution of the globular cluster
system through the evaluation of some time--dependent quantities, namely
the mass--dependent and global number densities:
$$n(M,\vec x,t)= \int_{\vec v} F(M,\vec x,\vec v,t) d^3\vec v,\eqno (8a)$$
$$n(\vec x,t)= \int_0^\infty n(M,\vec x,t)dM,\eqno (8b)$$
total number:
$$N_{gc}(t)=\int_{\vec x} n(\vec x,t)d^3\vec x,\eqno (8c)$$
total mass:
$$M_{gc}(t)=\int_{\vec x} n(M,\vec x,t)Md^3\vec xdM,\eqno (8d)$$
mass function:
$$\Psi(M,t)=\int_{\vec v}\int_{\vec x} F(M,\vec x,\vec v,t)
d^3\vec xd^3\vec v.\eqno (8e)$$
\par For the sake of simplicity, but without losing too much in "generality",
we make the following assumptions:
\par
i) the initial distribution is factorizable in its mass dependence (i.e.
the initial mass distribution is the same in any point and velocity);
ii) the time evolution of the distribution function is due to
frictional and tidal depopulation only, i.e. $\dul F\equiv (\dul F)_{df}+
(\dul F)_{tid}=-F/\tau_{df}-F/\tau_{tid}$;
iii) we consider only radial space dependences, thus neglecting second
order geometric effects.
\par As a consequence of i) to iii) we can write:
$$F(M,\vec x,\vec v,t)=F_0(M,\vec x,\vec v)e^{-t/\tau_{tid}}e^{-t/\tau_{df}}=
f_0(r,\vec v)\Psi_0(M)e^{-t/\tau_{tid}}e^{-t/\tau_{df}}.\eqno (9)$$
The overall evolution of the system  relies thus on a suitable choice of
the initial phase--space density (hereafter referred as DF) and of the IMF.
In the above approximation, the mass loss rates of the globular cluster system
due to dynamical friction and tidal disruption are, respectively:
$$\left[\dul M_{gc}(t)\right]_{df}=
-\int_0^\infty \int_0^\infty \int_{\vec v}
{1\over \tau_{df}} F(M,\vec x,\vec v,t)Md^3\vec v4\pi r^2drdM, \eqno(10a) $$
$$\left[\dul M_{gc}(t)\right]_{tid}=
-\int_0^\infty \int_0^\infty \int_{\vec v}
{1\over \tau_{tid}} F(M,\vec x,\vec v,t)Md^3\vec v4\pi r^2drdM.\eqno (10b)$$
In the following we follow the galactic nucleus
growth rate as a fraction $0\leq \alpha\leq 1$ of frictionally decayed mass
$$\dul M_n(t)=\alpha \left|\left[\dul M_{gc}(t)\right]_{df}\right|.\eqno
(10c)$$
To abbreviate, hereafter $\left[\dul M_{gc}(t)\right]_{df}$
and $\left[\dul M_{gc}(t)\right]_{tid}$ will be referred simply
as $\dul M_{df}$ and $\dul M_{tid}$.
In this paper $\alpha$ is considered as a free parameter.
The system evolution (and in particular that of the nucleus mass)
is obtained by integration of
the system of (10a),(10b),(10c) with
the initial conditions $M_{df}(0)=M_{tid}(0)=0$, $M_n(0)= M_{n0}$.
In the numerical models presented in Sect.4
I always take $M_{n0}=0$. In line of principle, the dynamical friction
decay time $\tau_{df}$ in equations (9) and (10) as given in Appendix A
is an overestimate, because it is computed in the density distribution
of the Schwarzschild model and so it does not take into account
the star density enhancement in the central galactic region due to
the cluster orbital decay itself. Anyway, PCV showed how the
overwelming contribution to dynamical friction is rather impulsive during
the galactic core crossing and linear in the background density,
such suggesting as suitable a corrective scaling of $\tau_{df}$
with a factor $\rho_c(0)/\rho_c(t)=1+[M_{gc}(t)-M_{gc}(0)-M_n(t)]/M_c(0)$.
The autoconsistency of our model in itself would require that
this factor is not far from unity: indeed in all our computations
it is found never to exceed 1.16, with a variation of the model's
quantities less than $5\%$.
\medskip\centerline
{\it 3.1 The Choice of Distribution Function and IMF}
\medskip\noindent
As cluster IMF, in this paper we shall
assume truncated power laws:

$$\Psi_0(M)=\cases{0 & $M<M_{min}$;\cr
	     M^{-s} & $M_{min}\leq M\leq M_{max}$;\cr
        	  0 & $M>M_{max}$;\cr}\eqno (11a)$$
 and gaussians:
$$\Psi_0(M)=\cases{0 & $M<M_{min}$;\cr
	          e^{-{1\over 2}{{(M-M_0)^2}\over \sigma_M^2}}
		     & $M_{min}\leq M\leq M_{max}$;\cr
		  0 & $M>M_{max}$;\cr}. \eqno (11b)$$
\par For the distribution function, we follow OBS  and consider
globular cluster DFs such to represent (i) an isotropic
distribution of globular clusters, (ii) a radially biased
situation (all clusters are on box orbits) and (iii) a situation
where the phase space is (in dependence on a parameter)
continuously populated along the tube/box frontier.
\par\noindent A distribution function which
accomplishes the request (i) is the isothermal DF:
$$f_0(E)=Ce^{{{\Phi_0-E}\over \sigma_0^2}} \eqno (12a).$$
Because we do not require that the initial density distribution of
globular clusters follows the underlying radial stellar distribution
(eq. (1)), C must be considered a normalization constant
and $\sigma_0$ a free parameter.
\par\noindent If all clusters were on box orbits (thing possible
if they are formed during a phase of strong radial collapse) a
realistic DF is:

$$f_b(E,L)=\cases{(L_c/L_{crit})^2f_0(E) & for $L<L_{crit}$;\cr
	           0  & for $L\geq L_{crit}$} \eqno (12b)$$
$L_c(E)$ is the circular angular momentum of a star of energy $E$ and
$L_{crit}(E)$ is the maximum angular action allowed to a box orbit of
energy $E$. We remind that $f_b(E,L)$ can be thought as generated
by a shift of points distributed in the action space according to $f_0(E)$
towards the $J_r$ axis for a fraction $1-L_{crit}/L_c$ at
constant $E$ (see OBS). This procedure requires $L_{crit}\leq L_c$.

\par\noindent As it was shown by OBS, a family of DFs which combine
a bias of box orbits with a continuous distribution across the frontier
between boxes and tubes (case (iii)) is defined by:
$$\eqalign{ f_k(E,L)=&
C {1-k^2r_c^2/(1+{1\over 2}k^2L^2/\sigma_0^2)\over
1+{1\over 2}k^2L^2/\sigma_0^2}e^{\Phi_0-E\over \sigma_0^2}\cr
& +C {k^2r_c^2\over 3\sqrt{3}(1+{1\over 6}k^2L^2/\sigma_0^2)^2}
e^{\Phi_0-E\over 3\sigma_0^2}}, \eqno (12c)$$
where $k$ is a sufficiently small constant.
\bigskip\centerline {4. RESULTS}
\medskip\centerline
{\it 4.1 The Nucleus Growth}
\medskip\noindent
We have computed various evolutionary models, trying to understand
their sensitivity to the various parameters in the IMF and in
the DF. For the sake of simplicity, we mainly refer and discuss here
models with power law IMF (eq. (11a)) and box orbit--biased DF
(eq. (12b)). These results are quite significant and appear
general enough; anyway, an analysis of the dependence of results on the IMF
and DF is given in Sect. 4.3.
\par Figures 4, 5, and 6 show the time evolution of $M_n$, $M_{tid}$,
--i.e. the nucleus accretion rate and the
rate (by mass) of globular cluster disruption--
their time derivatives, and the evolution of the total number and mass
of the globular cluster system.
\newline The family of globular clusters was populated
by $1000$ objects with
power law IMFs in the range from $10^4$ to $3\times 10^6$ M$_\odot$
characterized by three values of $s$ (0,2,3)
 and for three values of $\sigma_0$ in the  DF:
$0.5\sigma_{typ}$ (Fig. 4), $\sigma_{typ}$ (Fig. 5) ,$2\sigma_{typ}$
(Fig. 6). The value $\alpha=1$ is assumed, so $M_n=M_{df}$.
The model with $\sigma_0=\sigma_{typ}=290 Km/sec$, flat IMF and
$N_{gc0}=1000$ is considered in the following as the reference model.
\par The trend of a rapid initial
decay to the galactic center of most massive clusters, leading
to a fastly increasing $M_n$ followed by a rising efficiency
of the tidal shattering, is in common, but it depends quantitatively
in a strong way on $\sigma_0$. Indeed (see Fig. 4 c),
"cold" systems ($\sigma_0\lappreq 0.5\sigma_{typ}$) survive for a
relatively short time($\lappreq 3~Gyr$) even in the case of the steepest
(s=3) IMF. Of course, the steeper the mass function the smaller the
resulting quasi--stationary $M_n$ (see Figs. 4,5,6 b),
due to underabundance of massive clusters.
It is quite interesting to note that the amount of mass carried to
the galactic center is within a factor of two that tidally dispersed
(at least in the cases of $\sigma_0=0.5\sigma_{typ}$ and
$\sigma_0=\sigma_{typ}$).
The evolution of the number and total mass of surviving clusters
is given by Figures 4,5,6 panels c and f.
For $\sigma_0=\sigma_{typ}$, the system with flat
IMF halves its population in less than 1.5 Gyr, while the $s=2$ IMF
needs 10 Gyr (the total mass is reduced to half of its initial value in
1.2 and 7.1 Gyr, respectively).
Two regimes in the cluster depauperation are evident in the
case of low and intermediate velocity dispersion:
to an initial rapid decrease (lasting a time increasing
with the steepness of the IMF, up to about 5 Gyr for the
steepest IMF) follows a slow "evaporation".
Due to the mutual non linear feedback of $M_n$ in the equations
(10a,b,c), the global system evolution cannot be easily scaled
in the total mass of the cluster family M$_{gc0}$,
even though the time evolution of $N_{gc}$ and $M_{gc}$
is found to scale approximately as $N_{gc0}^{0.91}$ and
$M_{gc0}^{0.94}$, i.e. near to linearity (see Fig. 7).
Actually,
$\dul M_{df}$ shows initially a dependence on M$_{gc0}$ which is almost
linear, through the normalization constant of the DF, while
$\dul M_{tid}$ starts
almost quadratic in $M_{gc0}$. The ratio $\dul M_{df}/\dul M_{tid}$
results to be decreasing in time and initially scaling
approximately as $M_{gc0}^{-1}$ (Fig. 8).
This means that poorly populated globular cluster systems have an
evolution dominated by dynamical friction on a long time scale (even larger
than the age of ther galaxy), while a very abundant
globular cluster system rather would form rapidly a compact
central object wich later remains almost constant in mass and
acts as a destroyer of incoming lighter clusters.
Results of the reference model at varying $N_{gc0}$ show that the
fraction of the total
globular cluster mass gone within $20~Gyr$ to the center of the galaxy
is proportionally smaller at increasing $N_{gc0}$,
but only for a factor 3 at increasing of a factor 20
the available mass (from
$M_n(20~Gyr)=0.6M_{gc0}$ for $N_{gc0}=500$ to $M_n(20~Gyr)=0.2M_{gc0}$ for
$N_{gc0}=10000$).
Consequently the expected merged mass
is in any case $\approx 7$ times greater for the most
popolous system.
This simply reflects the physical intuition that, once fixed the
mass distribution shape and velocity distribution, the higher
the globular cluster number the higher the merged mass in a given
time.
To quantify it better, Fig. 9 shows the curve $t_{eq}(N_{gc0})$
which separates the evolution dominated
by dynamical friction (ages less than $t_{eq}$)
from that dominated by nucleus tidal interaction (ages greater than $t_{eq}$).
$t_{eq}$ ranges from 35 Myr for $N_{gc0}=10000$ to 2800 Myr for
$N_{gc0}=500$.
\par With regard to the role of a variation of $M_{min}$ and $M_{max}$
in the IMF (eq. (11a)) of the reference model, an increment of $M_{min}$
from $10^4~M_\odot$ to $5\times 10^4~M_\odot$ yields (at $t=20~Gyr$) to
almost the same value of $N_{gc}$, and to small increments of
$M_{gc}$, $M_n$ and $M_{tid}$ (15\%, 0.41\%, 2.3\%, as percentage
variations relative to the values of the reference model).
As expected, the sensitivity on $M_{max}$ is quite larger: changing
$M_{max}$ from $10^6~M_\odot$ to $3\times 10^6~M_\odot$ induces
precentage variations of $-69\%$, 52\%, 72\%, 66\% for $N_{gc}$, $M_{gc}$,
$M_n$ and $M_{tid}$, respectively.
\par To investigate the role of $\alpha$ in equations (10a,b,c) I computed
the evolution of models equivalent to the reference model with four different
values of $\alpha$, namely $10^{-4},10^{-3},10^{-2},10^{-1}$.
A case with $\alpha=0$ is studied in Sect. 5.
For $\alpha \leq 10^{-1}$
$M_n(t)$ results to scale approximately linearly in $\alpha$: this is
due to the small dependence of $\dul M_{df}(t)$ on the nucleus mass
caused by its exponential dependence on a factor
proportional to $-tM_n$. In the meantime, the time evolution of the
globular cluster total number and mass is not very sensitve
to $\alpha$ for $\alpha \leq 10^{-1}$; for instance the
number of survived clusters at 20 Gyr increases for just
a $10\%$ at reducing $\alpha$ from $10^{-1}$ to $10^{-4}$,
while it is greater ($14\%$) when $\alpha$ decreases from 1 to
$0.1$.
This is due to that for
not too massive nucleuses the contribution of $\dul M_{tid}$, which is
of course sensibly dependent on $M_{n}$, to the total globular
cluster depopulation is overwhelmed by $\dul M_{df}$.
\par
Finally, a good fitting formula to the final (at $t=20~Gyr$) value of
the nucleus mass is
$$M_n=0.0159\alpha F(s,\sigma/\sigma_{typ})G(M_{min},M_{max},s)
M_{gc0}^{0.74}M_{max}^{1.17}\eqno(13).$$
All masses are in $M_\odot$; the functions $F$ and $G$ in equation (13) are
$$F(s,\sigma/\sigma_{typ})=(6.348s^2-33.138s+42.73 )\left({\sigma\over
\sigma_{typ}}\right)^{-0.15}-5.243s^2+27.429s-35.5\eqno(13a),$$
and
$$G(M_{min},M_{max},s)=\cases{\left[ {{log_e(M_{max}/M_{min})}\over
{M_{max}-M_{min}}}\right]^{0.74} & $s=1$;\cr
\left[{{2-s}\over {1-s}} {{M_{max}^{1-s}-M_{min}^{1-s}}\over
{M_{max}^{2-s}-M_{min}^{2-s}}}\right]^{0.74} & $s\ne 1$;\cr}\eqno (13b).$$
The formula (13) is good within a $15\%$ error for $\alpha\leq 10^{-1}$,
$7.5\leq M_{gc0}/(10^8 M_\odot)\leq 30$, $0\leq s\leq 3$,
$0.5\leq \sigma/\sigma_{typ}\leq 2$, $10^4\leq M_{min}/M_\odot\leq 5\times
10^4$, $10^6\leq M_{max}/M_\odot\leq 3\times 10^6$.
\medskip\centerline
{\it 4.2  The Evolution of the Cluster Mass Function}
\medskip\noindent
The simultaneous effects of dynamical friction and tidal disruption
act mainly on opposite sides of the globular IMF, and so the
qualitative expectation is that the mass function evolves through
a progressive depopulation of higher masses followed by
that of light clusters, leading to a shape showing a peak
at some intermediate mass value which, in this case, is
a natural evolutive result of even an initial scale--free mass function
and not the preferred scale of cluster formation. This may mean
that the observed peak of the luminosity function of
globular clusters which is remarkably constant in various
galaxies (van den Bergh 1985) is a result of a similar
evolution of globular cluster systems in their parent galaxies
which, in their turn, should be similar, at least on the scales relevant to
the global evolution of the cluster sustems.
Figure 10 illustrates the time running of the MF for the
reference model and four
values of the nucleus conversion efficiency $\alpha$. As expected,
at decreasing $\alpha$ the low mass side of the MF tends to
be unaltered by tidal disruption while the effect of dynamical friction
progressively affects also low masses: this implies a reduction of the present
($\sim 15~Gyr$)
peak value of the mass function (from $\sim 3\times 10^5$ M$_\odot$
at $\alpha=1$, to $\sim 10^5$ M$_\odot$ at $\alpha=0.1$, to
$\sim 4\times 10^4$ M$_\odot$ at $\alpha=10^{-2}$)
down to disappearing of the low mass turn--over.
At decreasing $\alpha$ (and so the nucleus mass) the dynamical friction
inprint is clearer in determining a power law high mass tail with
a slope $-0.4$ (in the case of Figure 10).
If the IMF is not scale--free, as in the case of $s\neq 0$ power law and
gaussian MF, the evolution is of course depending on the preferred
scale. As an example, Figure 11 shows the evolution of a
gaussian IMF of the type of equation (11b), centered at $M_0 =10^6$ M$_\odot$,
$\sigma_M =M_0/2$.
\medskip\centerline
{\it 4.3 The Effects of Varying the DF}
\medskip\noindent
For the sake of comparison, we have performed computations
using an isotropic DF as given by equation (12a) and the distribution
function obtained letting $k=0.2/r_c$ in equation (12c).
The globular cluster family evolutions result in both cases to be very similar
to that of the box--biased DF. Indeed, the isotropic DF $f_0$
is significantly different from $f_b$ at $E>0.4$,
where the factor $(L_c/L_{crit})^2$ sharply increases.
Because $E=0.4$ is more than three times
$(3/2)\sigma_{typ}^2$ (this latter is the typical energy),
the majority of globular cluster have energies such
that the DFs are very similar: actually, at these energies ($\leq 0.4$)
the maximum angular momentum allowed by the general potential
to the $F_0$ and $f_b$ DFs is $L_{max}(E)=\max_{|\vec x|}
\sqrt{2[E-\Phi(\vec x)}]$
$<L_{crit}(E)$, so making undistinguishable in $L$ the two
distributions. For the same reasons also the DF $f_{k}$ gives
similar results for what is of interest here, thus confirming what found
by OBS in a different context: an anisotropic DF of the form given
by equation (13)  significantly differs from the isotropic only in the
velocity dispersion profiles.
\bigskip\centerline
{5. THE M 31 GALAXY}
\medskip\noindent
The isophotal behaviour of the bulge component of Andromeda was
shown (Stark 1977) to be triaxial with an estimate
1:0.625:0.5 for the  axial ratios,
i.e. the same of the galactic model used in this paper.
As a consequence, an application of our evolution
model to a simulation of some of the characteristics of M 31
is quite straightforward.
In addition, there is evidence for a presence of a compact mass
in its center of the order of few times $10^7$ M$_\odot$
(Dressler \& Richstone 1988; Kormendy 1988b); a recent paper
by Melia (1992) showed that radio and X observations of the central
M 31 region is compatible witha $10^7~M_\odot$ black hole.
\par The estimated number of globular cluster in M31 is
$\approx 375$ (Fusi--Pecci 1992), which is $2.5$ times that
of galactic globulars.  In the hypothesis that a nucleus of
mass $3\times 10^7$ M$_\odot$ is standing in the M31 center
since its formation, in order to reproduce that number
the dynamical--friction and nucleus tidal
effects on the system of globular clusters orbiting that galaxy
with a DF of the form ($12b$) with $\sigma_0=\sigma_{typ}$ requires an
initial population of $~1500$ (in the case of a flat IMF
cutted at $10^4$ and $3\times 10^6$ M$_\odot$) or of $~920$ clusters
(power law IMF with slope $-2$). The number of clusters is thus reduced to
$25\%$ and $41\%$, respectively, from the initial.
Table 2 shows some characteristics
of the models with the two different IMFs.
In the case of the decreasing IMF,
the global efficiency of the tidal "erosion" of globular system results
similar to that of dynamical friction while the presence of an
initial high number of massive clusters (as it is with the flat IMF)
resulted into that only $~4\%$ of the initial system mass has been affected
by tidal disruption within an Hubble time. For the sake of simplicity
we have not considered here the possible contribution to the nucleus
mass by decayed globulars, i.e. we let $\alpha=0$.
\medskip\centerline
{\it 5.1 Dynamical Friction Orbital Decay as Supporting Violent Galactic
Activity}
\medskip\noindent
The short halving time of $\dul M_{df}$ in the M 31 model
(less than $4.5\times 10^8 yr$,
which is incidentally similar to the Eddington time scale)
and its initial high value suggests that the rapid release
of power due to fall of mass on a compact central object
caused by dynamical friction
may have been significant for the initial galactic activity.
To check this, we computed the maximum luminosity
$L_m\equiv \dul M_{df}c^2$ and its ratio to
the Eddington luminosity as function of time, for the two models
of Table 2. The curves are shown in Figure 12.
The true power emitted should have been $L=\xi L_m$, with
$\xi \sim 0.1$.
A regime of quasi--constant high luminosity lasts for the first
$10^7$ yr, to which a regime $L_m\propto t^{-{3\over 2}}$ follows.
Note (Fig. 12b) that the case of decreasing IMF corresponds to
a $L_m$ which is always sub--Eddington, while the case of flat
IMF could have been super-Eddington (assuming $\xi =0.1$,
$L$ has been super--Eddington for the first $0.8$ Gyr).
\par More specific conclusions related to an observational
comparison with the M31 globular clusters and a more straightforward
analysis of the outlined possible origin of galactic activity require
a deeper and more refined study to be accomplished. Anyway,
the previous considerations seem to indicate that
also a galaxy like M31 --which now appears normal and quiet-- could have
been active in the past; the total bolometric luminosity
supported by the dynamical friction mass feeding of
a pre-existing nucleus and the relative time
scale are shown to be in the acceptable range of observed AGNs.
\bigskip\centerline
{6. Conclusions}
\medskip\noindent
In this paper we investigated the importance of the two phenomena
of dynamical friction and nucleus tidal disruption
(particularly relevant in triaxial galaxies) in determining
the evolution of a system of satellites orbiting in the potential
of a triaxial galaxy, here modeled with the Schwarzschild (1979)
ellipsoid.
As expected, the two phenomena are competitive and the first
important result
is that they actually determine the global evolution of the
globular cluster system in the galaxy, being active on time scales
which are significantly shorter of the age of the galaxy. Dynamical
friction is overhelming whenever the mass of the compact nucleus
is not too large. The value of this mass depends of course on
the global characteristics of the globular cluster systems,
in particular on their energy and angular momentum distribution.
Typically, taking as individual globular cluster mass
$10^5$, $10^6$, and $10^7$ M$_\odot$, the value of
the mass needed by the nucleus to shatter clusters faster than
frictional braking is $\approx 5\times 10^6$ M$_\odot$,
$\approx 2\times 10^8$ M$_\odot$
and $\approx 5\times 10^9$ M$_\odot$ respectively. This means that
if a galaxy has formed a central nucleus at the epoch of its birth,
if it was initially not too massive
its mass could have been significantly grown up
due to cannibalism of orbitally decayed globular clusters.
\par The study of how and with what efficiency $\alpha$
($\dul M_n=\alpha \dul M_{df}$) a nucleus can actually form and
grow was out of the
purposes of this work and deserves careful investigation.
Here we have checked that the quantity of stellar mass carried to the
galactic centre in form of braked globular clusters is high enough
to grow up a compact object of a size and on a time scale which
depends (at fixed efficiency $\alpha$)
sensitively on the velocity dispersion of the globular
cluster system and on its mass function. Assuming the limiting
value $\alpha=1$ the most massive nucleus grew up to $\sim10^9 M_\odot$
in $\sim 3\times 10^7 yr$.
\par The fast decay of most massive globulars
should have been a possible cause of a violent transient activity
in galaxies. As an example, a model of M 31 as having {\it ab initio}
a nucleus of $\sim 3\times 10^7 M_\odot$ shows that a
high gravitational luminosity
($5.6\times 10^{10}\leq \dul Mc^2/L_\odot\leq 2.8\times 10^{12}$)
can be supported for about $400~Myr$. This could mean that even quite
normal present galaxies have been active in the past.
\newline As a final remark, in this paper it was shown that
the 2 phenomena studied tend contemporarily to deplete the
opposite sides of the IMF, in such a way that even a scale--free
IMF evolves to show a definite maximum, whose value is so
not representative of the characteristic scale of fragmentation
but rather depends mostly on the initial range of masses involved.
\bigskip
\centerline {\bf Appendix A}
\oneskip
A good fit to $\tau_{df}$ for box orbits
in dependence on the energy per unit mass
$E\equiv {1\over 2} v^2+\Phi(\vec x)$ and on the mass $M$ of the cluster
is given (for $0<E<1$) by:
$$\tau_{dfb}(E,M)= \left({{10^6\over M}}\right)
{7.5\times 10^8\over (1-E)^2},\eqno (A1)$$
if $M$ is given in solar masses $\tau_{dfb}$ is in years.
Decay times of clusters on loop orbits can be scaled by that of
box orbit of same energy via a function of energy and azimuthal action $J$
(assumed as $J=0$ for the box orbit):
$$\tau_{df}(E,M,J)=10^{g(E,J)}\tau_{dfb}(E,M)\eqno (A2)$$
where:
$$g(E,J)=1.564E^{5/2}\sqrt{\left({J\over J_c}\right)^
{1\over E+0.1}} +0.3E \eqno(A3)$$
$J_c$ is the azimuthal action of the quasi--circular orbit of energy $E$.
\newline The average (over $E$ and $J$) error is $\sim 10\%$.
\vfill\eject
{}~
\vfill\eject
{}~
\vfill\eject
\centerline{REFERENCES}
\oneskip
\ref Binney, J. \& Tremaine, S., 1987, Galactic Dynamics,
(Princeton: Princeton University Press)
\ref Chandrasekhar, S., 1942, Principles of Stellar Dynamics,
(New York: Dover)
\ref de Zeeuw, T. \& Merritt, D., 1983, ApJ, 267, 571
\ref Dressler, A. \& Richstone, D.O., 1988, ApJ, 324, 701
\ref Fusi--Pecci, F., 1992, XIth S. Cruz Summer Workshop,
eds. J. Brodie \& G. Smith, ASP Conf. Ser., in press
\ref Gerhard, O.E. \& Binney, J., 1985, MNRAS, 216, 467
\ref Kormendy, J., 1988a, ApJ, 325, 128
\ref Kormendy, J., 1988b, ApJ, 335, 40
\ref Kormendy, J. \& Richstone, P., 1992,ApJ, 393, 559
\ref Melia, F., 1992, ApJ, 398, L95
\ref Merritt, D., 1980, ApJS, 43, 435
\ref Ostriker, J.P., Binney, J. \& Saha, P., 1989, MNRAS, 241, 849
\ref Pesce, E. Capuzzo--Dolcetta, R. \& Vietri, M., 1992, MNRAS, 254, 466
\ref Schwarzschild, M., 1979, ApJ, 232, 487
\ref Stark, A.A., 1977, ApJ, 213, 368
\ref Surdin, V.G. \& Charikov, A.V., 1977, Sov. Astron., 21, n.1, 12
\ref Tonry, J.L., 1984, ApJ, 283, L27
\ref Tonry, J.L., 1987, ApJ, 322, 632
\ref Tremaine, S., Ostriker, J.P. \& Spitzer, L., 1975, ApJ, 196, 407
\ref van den Bergh, S., 1969, ApJS, 19, 145
\ref van den Bergh, S., 1985, ApJ, 297, 361
\ref van den Bergh, S., 1991, PASP, 103, 1053
\ref Webbink, R.F., 1985, in IAU Symp. 113, Dynamics of Star Clusters,
eds. J. Goodman \& P. Hut,(Dordrecht:Reidel), 541
\vfill\eject
\centerline {\bf Figure Captions}
\bigskip
\noindent {\bf Figure 1}
\newline
The ratios between the energy  decay times of the box
orbit of energy E to that of the corresponding quasi--circular orbit
(solid line) and to that of the most elongated loop (dashed line).
Dotted line is the ratio between the energy total
decay time and the energy halving time, $\tau_{1/2}/\tau_{df}$,
of the box orbit of energy E.
\bigskip\noindent {\bf Figure 2}
\newline
The (approximate) relation between the centrally merged cluster mass
and the cluster initial mass (in solar units), for three orbital
energies: $E=\bar E=0.6$ (solid line), $E=\bar E/2= 0.3$ (dashed
line); $E=\bar E/4=0.15$ (dotted line);
The curves are labelled with the time needed to merge the mass corresponding to
$10^6 M_\odot$ clusters.
\bigskip\noindent {\bf Figure 3}
\newline
Solid line is the maximum mass, $M_{l,df}$, and dashed line
the minimum, $M_{l,tid}$, allowed by dynamical friction and
tidal disruption, respectively, for cluster on box orbits of
energy E, evaluated at the ages 1, 5, 10, 15 Gyr. The 1 Gyr
and the 15 Gyr curves are labeled. The shaded area is the
area permitted at present day to globulars.
The two arrows
correspond to the heaviest ($\omega Cen$) and lightest (AM 4)
globulars in the Webbink (1985) compilation.
\bigskip\noindent {\bf Figure 4}
\newline
Running vs. time of several model variables, for
the box--biased DF with
$\sigma_0=0.5\sigma_{typ}$ and power law IMF in the
$10^4\div 3\times 10^6 M_\odot$ interval with
exponent $s=0$ (solid line), $s=2$ (dotted line), and
$s=3$ (dashed line). Masses are in $M_\odot$ and
time derivatives in $M_\odot yr^{-1}$.
\bigskip\noindent {\bf Figure 5}
\newline
As in Figure 4 but with $\sigma_0=\sigma_{typ}$.
\bigskip\noindent {\bf Figure 6}
\newline
As in Figure 4 but with $\sigma_0=2\sigma_{typ}$.
\bigskip\noindent {\bf Figure 7}
\newline
The behaviour in the two first billion years of the globular
cluster number (panel a) and total mass (panel b)
scaled to their initial
values for four different initial abundance of the sample
in the reference model: $N_{gc0}$=500 (solid line);
1000 (dotted line); 2000 (short--dashed line); 10,000 (long--dashed
line).
\bigskip\noindent {\bf Figure 8}
\newline
The logarithmic ratios between $\dul M_{df}$ and $\dul M_{tid}$ vs.
age, for the reference model with different initial number of cluster s.
Line caption as in Figure 7.
\bigskip\noindent {\bf Figure 9}
\newline
In function of the initial abundance ($N_{gc0}$) of the globular clusters
sample in the reference model, the curve in this figure indicates
the age up to which dynamical friction overwhelms tidal disruption.
\bigskip\noindent {\bf Figure 10}
\newline
The behaviours of the globular cluster system mass
function at the ages (in Gyr) $0, 1.24, 5.41, 14.8, 20$ from top to bottom.
The model is the reference one.
Each panel is labeled with
the corresponding value of the efficiency parameter $\alpha$.
Abscissa and ordinate
(labeled only in the lower left panel) are in solar units.
\bigskip\noindent {\bf Figure 11}
\newline
The behaviours of the globular mass functions at the same ages
of Figure 10, but for a gaussian IMF centered at $10^6 M_\odot$,
with standard deviation $5\times 10^5 M_\odot$ and cutted at
$10^4$ and $3\times 10^6 M_\odot$.
\bigskip\noindent {\bf Figure 12}
\newline
The time behaviour of the maximum luminosity
$\L_m$ defined in Sect. 5.1 in solar units (panel a) and
in Eddington luminosity units (panel b). Solid line refers to the M 31
model with flat IMF; dotted line to the $s=2$ IMF (see Table 2).
\bye